\documentclass[conference]{IEEEtran}
\IEEEoverridecommandlockouts
\usepackage{cite}
\usepackage{amsmath,amssymb,amsfonts}
\usepackage{algorithmic}
\usepackage{graphicx}
\usepackage{textcomp}
\usepackage{xcolor}
\def\BibTeX{{\rm B\kern-.05em{\sc i\kern-.025em b}\kern-.08em
    T\kern-.1667em\lower.7ex\hbox{E}\kern-.125emX}}

\usepackage{algorithm}
\usepackage{array}
\usepackage[caption=false,font=normalsize,labelfont=sf,textfont=sf]{subfig}
\usepackage{stfloats}
\usepackage{url}
\usepackage{verbatim}
\usepackage{cite}
\usepackage{enumerate}
\usepackage{caption}
\usepackage{enumitem}
\usepackage{tabularx,booktabs,textcomp}
\usepackage{amsthm}

\usepackage{amsmath,amssymb,amsthm}

\begin{document}

\title{Identification of Forced Oscillation Sources in Wind Farms using E-SINDy}

\author{\IEEEauthorblockN{K. Victor Sam Moses Babu}
\IEEEauthorblockA{\textit{Dept. of EEE} \\
\textit{BITS Pilani Hyderabad Campus}\\
Hyderabad, India \\
victor.babu@in.abb.com}
\and
\IEEEauthorblockN{Dr. Pratyush Chakraborty}
\IEEEauthorblockA{\textit{Dept. of EEE} \\
\textit{BITS Pilani Hyderabad Campus}\\
Hyderabad, India \\
pchakraborty@hyderabad.bits-pilani.ac.in}
\and
\IEEEauthorblockN{Dr. Mayukha Pal*}
\IEEEauthorblockA{\textit{ABB Ability Innovation Center}\\
\textit{Asea Brown Boveri}\\
Hyderabad, India \\
*mayukha.pal@in.abb.com}
}

\maketitle

\begin{abstract}
The rapid growth of wind power generation has led to increased interest in understanding and mitigating the adverse effects of wind turbine wakes and forced oscillations in wind farms. In this paper, we model a wind farm consisting of three wind turbines connected to a distribution system. Forced oscillations due to wind shear and tower shadow are injected into the system. If these oscillations are unchecked, they could pose a severe threat to the operation of the system and damage to the equipment. Identifying the source and frequency of forced oscillations in wind farms from measurement data is challenging. Thus, we propose a data-driven approach that discovers the underlying equations governing a nonlinear dynamical system from measured data using the Ensemble-Sparse Identification of Nonlinear Dynamics (E-SINDy) method. The results suggest that E-SINDy is a valuable tool for identifying sources of forced oscillations in wind farms and could facilitate the development of suitable control strategies to mitigate their negative impacts.
\end{abstract}


\section{Introduction}
\label{section:Introduction}

The use of wind power as a renewable energy source has increased significantly in recent years, but it also poses new challenges due to its intermittent and variable nature \cite{trends, DD_arxiv}. Forced oscillations in wind farms are one of the most significant challenges, as they cause decreased power output and increased loads on turbines and support structures \cite{Wang2016}. These oscillations are triggered by various sources, such as wind shear, wind turbulence, and upstream turbine wakes, which could excite natural oscillations in the wind farm system and cause large amplitude vibrations that may damage equipment and reduce power output \cite{Lackner2013}. Identifying the sources of forced oscillations is critical to developing effective control strategies to mitigate their negative effects. However, the complexity and nonlinearity of wind farm systems make this identification challenging \cite{victor_wind}.

Physics-informed machine learning methodologies are used in identifying the dynamics of complex systems; one such method is the combination of the Reynolds-averaged Navier-Stokes (RANS) equations with machine learning methods. The RANS equations are a set of partial differential equations that describe the averaged behavior of fluid flow in a turbulent system. By combining the RANS equations with machine learning methods, it is possible to identify the underlying nonlinear dynamics of a wind farm system from limited data \cite{Peherstorfer2020, Zhang2021}. Another method that has been proposed is the use of physics-constrained neural networks (PCNNs) \cite{Meng2020}. PCNNs combine traditional neural networks with partial differential equations provide a description of the underlying physical processes in systems. By constraining the neural network to satisfy the physical equations, PCNNs may reduce the amount of data required for accurate model identification and provide a physically interpretable model of the system's behavior. PCNNs have been successfully applied to identify the nonlinear dynamics of various physical systems, including fluid flow and chemical reactions \cite{Raissi2018}. In the context of wind energy, PCNNs have been used to model the fluid dynamics of a wind turbine blade \cite{Pawar2021}. Machine learning techniques have also been applied in recent years to improve wind turbine and wind farm performance by predicting power output, detecting faults, and enhancing wind energy production \cite{Jin2019, Liu2019}. 

However, machine learning methods typically do not provide a physical understanding of the underlying dynamics \cite{Pal_Dynamics} of wind turbines or wind farms. As a result, there has been growing interest in using data-driven methods to discover the governing equations of physical systems from data, a task known as equation discovery or system identification \cite{Sapsis2018}. One widely used data-driven method is sparse identification of nonlinear dynamics (SINDy) \cite{Brunton2016}, which has been successfully applied to various dynamical systems. Researchers have successfully used SINDy to discover the dynamics of various physical systems, including fluid flow and chemical reactions \cite{Rudy2017}. However, the traditional SINDy method assumes that the system's dynamics are linear or low-order nonlinear, which may not hold for wind farm systems with complex interactions between turbines and the atmosphere \cite{Loiseau2021}. To overcome this limitation, an extension of SINDy known as ensemble-SINDy (E-SINDy) has been proposed \cite{ensemble}. E-SINDy identifies the underlying nonlinear dynamics of a system by using a collection of data sets that span the full range of the system's dynamics. Researchers have shown that E-SINDy is effective in identifying the nonlinear dynamics of various physical systems, including a fluid mixing system \cite{Chen2020} and a population of oscillators \cite{Jin2019}.

The E-SINDy method is proposed to identify the underlying nonlinear dynamics of wind farms from time-series data. By providing physically interpretable models of wind farm behavior, our work aims to enable more efficient operation and maintenance of wind turbines and wind farms, leading to increased wind energy production and reduced costs.

This work makes the following main contributions:

\begin{enumerate}
    \item Building the governing equations of the wind turbines from measurement data using the E-SINDy method. 
    \item Identification of the source and frequency of the oscillations from the governing equations built by the proposed methodology.
    \item Evaluating the effectiveness of the proposed methodology for different cases of forced oscillations in wind farms and its potential to improve wind farm operation and maintenance.   
\end{enumerate}

In the next sections, we will present the methods used in Section \ref{sec:methodology}. Then, we will present the results of the simulation study with real-world data in Section \ref{sec:Simulation}. Finally, we will draw our conclusions in Section \ref{section:Conclusion}.

\section{Methodology}
\label{sec:methodology}

\subsection{Mathematical Model for analyzing forced oscillations}

For wind turbines with induction generators (IGs), the rotor dynamics may  be modeled by modifying the swing equations to account for the differences in torque-speed characteristics and the slip between the rotor and the stator, represented as follows:
\begin{flalign}
\begin{split}
\dot{\delta} &= \omega - \omega_s \\
\frac{d}{dt}(J \omega) &= T_m - T_e - D(\omega - \omega_s) + T_f - T_L
\end{split}
\label{eq:induction_generator_stochastic_dynamics}
\end{flalign}

where  $\delta$ is the rotor angle,  $\omega$ is the rotor speed, $\omega_s$ is the synchronous speed,     $J$ is the moment of inertia of the rotor,  $T_m$ is the mechanical torque (from the wind turbine), $T_e$ is the electromagnetic torque (generated by the induction generator),    $D$ is the damping coefficient. $T_L$ represents the stochastic load torque that could be modeled as a random process, such as a Gaussian white noise process or an Ornstein-Uhlenbeck process, depending on the nature of the load variation. $T_f$ is the additional torque term representing the external input that causes forced oscillations, such as wind shear, tower shadow, or grid disturbances. This torque term may be modeled as a function of time, rotor angle, or other relevant variables. If multiple sinusoidal components with different frequencies, amplitudes, and phase shifts contribute to the forced oscillation torque, we could represent the overall torque as a sum of multiple sinusoidal terms. In this case, the Fourier series representation of the forced oscillation torque would be:
\begin{multline}
   T_f(t) = \frac{a_0}{2} + \sum_{n=1}^N \Bigl[a_n \cos(\omega_{f_n} t) + b_n  \sin(\omega_{f_n} t)\Bigr] 
\end{multline}
where $N$ is the number of harmonic components, $a_0$ is the constant term (representing the average value of the function), $a_n$ and $b_n$ are the coefficients of the cosine and sine terms of the $n$-th harmonic component, respectively, and $f_n$ is the frequency of the $n$-th harmonic component. The governing equations of the wind turbine under small perturbation can be rewritten as below, 
\begin{equation}
\begin{bmatrix}
\Delta \dot{\delta} \\
\Delta \dot{\omega}
\end{bmatrix} = 
\begin{bmatrix}
\Delta (\omega - \omega_s) \\
\Delta \left(\frac{1}{J} (T_m - T_e - D(\omega - \omega_s) - T_L)\right)
\end{bmatrix}
+
\begin{bmatrix}
0 \\
\frac{T_f}{J}
\end{bmatrix}
\label{eq:induction_generator_gov_eq}
\end{equation}

\begin{figure}[t]
  \centering
  \includegraphics[width=3.4in]{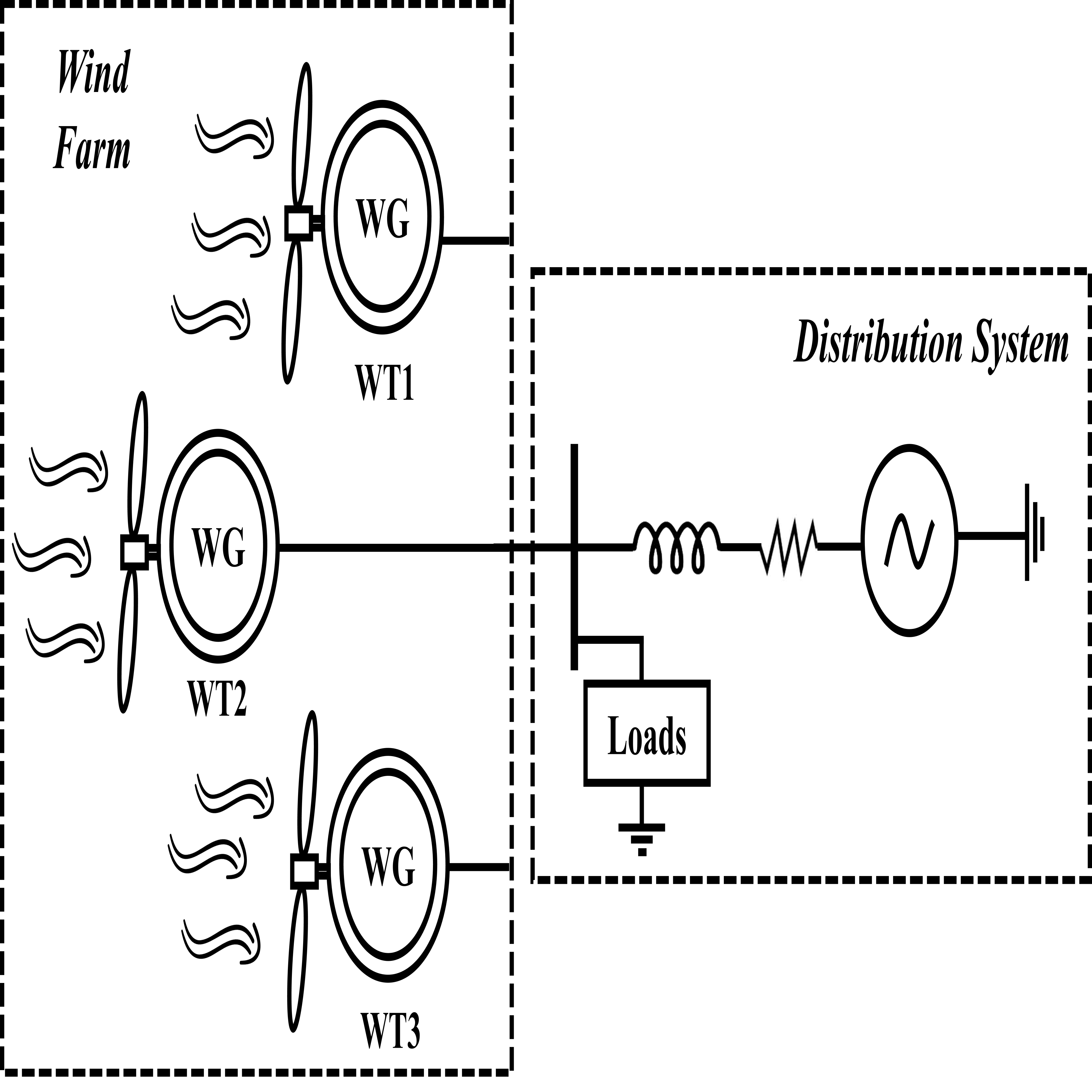}
  \caption{Schematic of wind farm connected to the distribution system.}
 \label{fig:schematic}
\end{figure}

\subsection{Mathematical model for the application of E-SINDy}
\begin{figure}[t]
   \centering
   \includegraphics[width=3.4in]{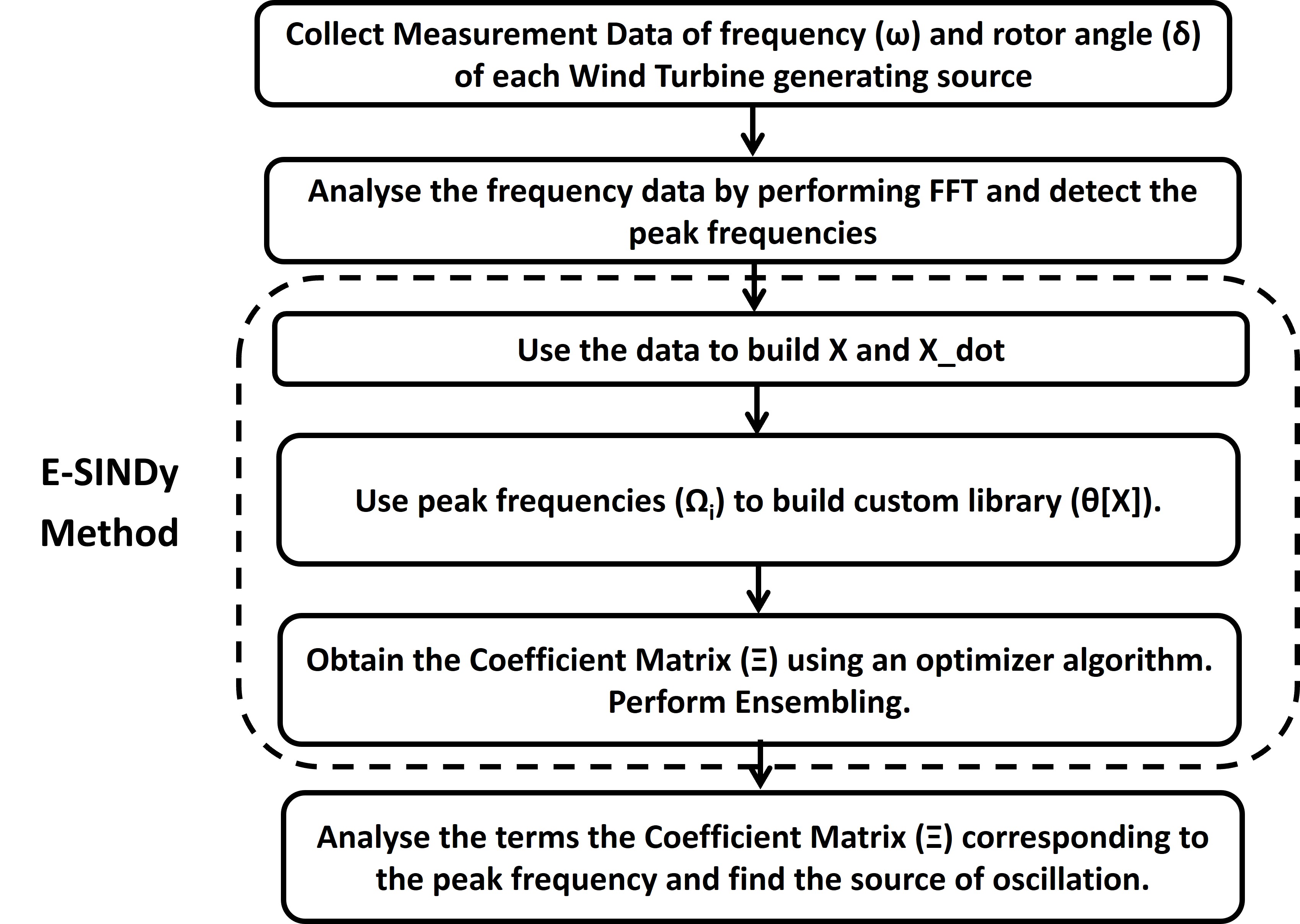}
   \caption{Steps for the proposed algorithm.}
   \label{fig:flowchart}
\end{figure}

In this work, we apply the extended sparse identification of nonlinear dynamics (E-SINDy) algorithm \cite{Rudy2017} to identify the underlying nonlinear dynamics of wind farm systems from high-fidelity simulation data. A schematic diagram of a wind farm with three wind turbines connected to a distribution system is shown in Fig. \ref{fig:schematic}. A step-by-step flow of the proposed method is provided in Fig. \ref{fig:flowchart}; starting with data collection to the identification of forced oscillation at the source of the wind turbine. We consider the general form of a nonlinear dynamical system,
\begin{equation}
\dot{\mathbf{x}} = \mathbf{f}(\mathbf{x})
\label{eq:nonlinear_dynamics}
\end{equation}

where $\mathbf{x}$ is the state vector and $\mathbf{f}(\mathbf{x})$ is the vector of functions that describe the dynamics of the system.
The E-SINDy algorithm aims to identify the nonlinear functions $\mathbf{f}(\mathbf{x})$ by leveraging the sparsity-promoting properties of the LASSO regression algorithm \cite{Tibshirani1996} and the ensemble approach \cite{Montavon2012}. The algorithm involves generating a library of candidate functions, which are nonlinear combinations of the state variables up to a certain polynomial degree. The library is then used to construct a matrix of measurements, $\mathbf{X}$, which contains measurements of the state variables at each time step.

The E-SINDy algorithm seeks to solve the following optimization problem:
\begin{equation}
\min_{\boldsymbol{\theta}} \left| \mathbf{X} - \boldsymbol{\Theta}(\mathbf{X}) \right|_F^2 + \lambda \left| \boldsymbol{\theta} \right|_1
\label{eq:lasso}
\end{equation}

where $\boldsymbol{\theta}$ is a vector of coefficients that determine the nonlinear terms in the system dynamics, $\boldsymbol{\Theta}(\mathbf{X})$ is a matrix of candidate functions evaluated at each time step, and $\lambda$ is a regularization parameter that controls the sparsity of the solution. The solution to \eqref{eq:lasso} provides a set of coefficients that could be used to construct the nonlinear functions $\mathbf{f}(\mathbf{x})$. The E-SINDy algorithm could also identify the structure of the system dynamics, including the number and types of nonlinear terms and the coupling between state variables. The steps are detailed in Fig. \ref{fig:flowchart}

The matrix $X^T$ is provided as a set of measurements at different time instances $t_1$ through $t_m$, along with its time-derivative matrix $\dot{X}$. These matrices could either be measured or computed using numerical methods. The time derivative matrix $\dot{X}$ may be expressed as a sum of linear combinations of columns from a feature library denoted by $\Theta$,

\noindent The measurement matrix $X$ is defined as follows, 
\begin{equation}
\begin{aligned}
X &=
\left[\begin{matrix}
  \Delta \delta_1(t_1) &\dots &\Delta \delta_r(t_1)  \\
 \vdots               &\ddots &\vdots                \\
\Delta \delta_1(t_m) &\dots &\Delta \delta_r(t_m)
\end{matrix}\right.\\
&\qquad\qquad
\left.\begin{matrix}
  {}&\Delta \omega_1(t_1) &\dots &\Delta \omega_r(t_1) \\
  {} & \vdots              &\ddots &\vdots  \\
  {} &\Delta \omega_1(t_m) &\dots &\Delta \omega_r(t_m)
\end{matrix}\right]
\end{aligned}
\end{equation}
\begin{align}
\dot{X}=
\begin{bmatrix}
\Delta \dot{\delta}_1(t_1) &\dots &\Delta \dot{\omega}_r(t_1) \\
\vdots &\ddots &\vdots \\
\Delta \dot{\delta}_1(t_m) &\dots &\Delta \dot{\omega}_r(t_m)
\end{bmatrix}
\end{align}

In the E-SINDy method, multiple bootstrapped samples of the original data matrix $X$ are generated by randomly selecting data points with replacements. The SINDy algorithm is applied for each bootstrapped dataset to identify a candidate model in the form of $\dot{X} = \Theta(X) \Xi$. This ensemble of candidate models addresses the sensitivity of the SINDy algorithm to noise and data variability. Finally, the best-performing model is selected based on a suitable model selection criterion, such as cross-validation or other statistical measures, providing a more robust and accurate representation of the underlying system dynamics.
For the selection of the feature library, the terms are chosen naturally, 

\begin{equation}
\begin{aligned}
\Theta(X) &=
\left[\begin{matrix}
  | &|               &\dots    &|               &|               &\dots  &|               &\dots \\
  1 &\Delta \delta_1 &\dots    &\Delta \delta_r &\Delta \omega_1 &\dots  &\Delta \omega_r &\dots \\
  | &|               &\dots    &|               &|               &\dots  &|               &\dots 
\end{matrix}\right.\\
&\qquad\qquad
\left.\begin{matrix}
  {} &|                   &|                   &\dots \\
  {} &\sin(\omega_{F_i t}) &\cos(\omega_{F_i t}) &\dots\\
  {} &|                   &|                   &\dots 
\end{matrix}\right]
\end{aligned}
\end{equation}

The coefficient matrix $\Xi$ consists of coefficient terms $\xi_1, \xi_2 \dots , \xi_n$. For the considered system, $\xi_{a}$ and $\xi_{b}$ correspond to the coefficients of the two terms in equation (\ref{eq:induction_generator_gov_eq}). The elements of forced oscillation at the wind generator of frequency $\omega_{F_i}$ are represented by,

\begin{align}
\xi_{b}^T=
\begin{bmatrix}
0 &\dots &a_{11} &\dots &a_{1r}\\
\vdots &\dots &b_{11} &\dots &b_{1r}\\
\vdots &\ddots &\vdots &\ddots &\vdots\\
\vdots &\dots &a_{n1} &\dots &a_{nr}\\
0 &\dots &b_{n1} &\dots &n_{nr}\\
\end{bmatrix}
\end{align}

The values of $a_{ij}$ and $b_{ij}$, where $i$ belongs to the set of integers from $1$ to $n$, and $j$ belongs to the set of integers from $1$ to $r$, represent the amplitude of input forced oscillation at the $j^{th}$ generator's frequency $\omega_{F_i}$. If the input forced oscillations are present in $X$, then the elements of $\xi_{b}$ will correspond to the forced oscillations due to the wind turbine source.

\section{Simulation Model \& Results}
\label{sec:Simulation}

We modeled a wind farm consisting of three wind turbines WT1, WT2, and WT3. The wind farm is connected to a distribution system. The simulation model is developed in MATLAB/SIMULINK. The frequency range for low-order oscillations from wind turbines is 0.388 to 0.775 Hz \cite{su_Wind}. Thus, we analyze the wind farm for three different cases:
\begin{enumerate}
    \item Case 1: 0.71 Hz oscillation in WT1.
    \item Case 2: 0.71 Hz oscillation in WT1 and 0.53 Hz oscillation in WT3.
    \item Case 3: 0.71 Hz oscillation in WT1, 0.61 Hz oscillation in WT2 and 0.53  Hz oscillation in WT3.
\end{enumerate}

\begin{figure*}
   \centering
   \includegraphics[width=6.5in]{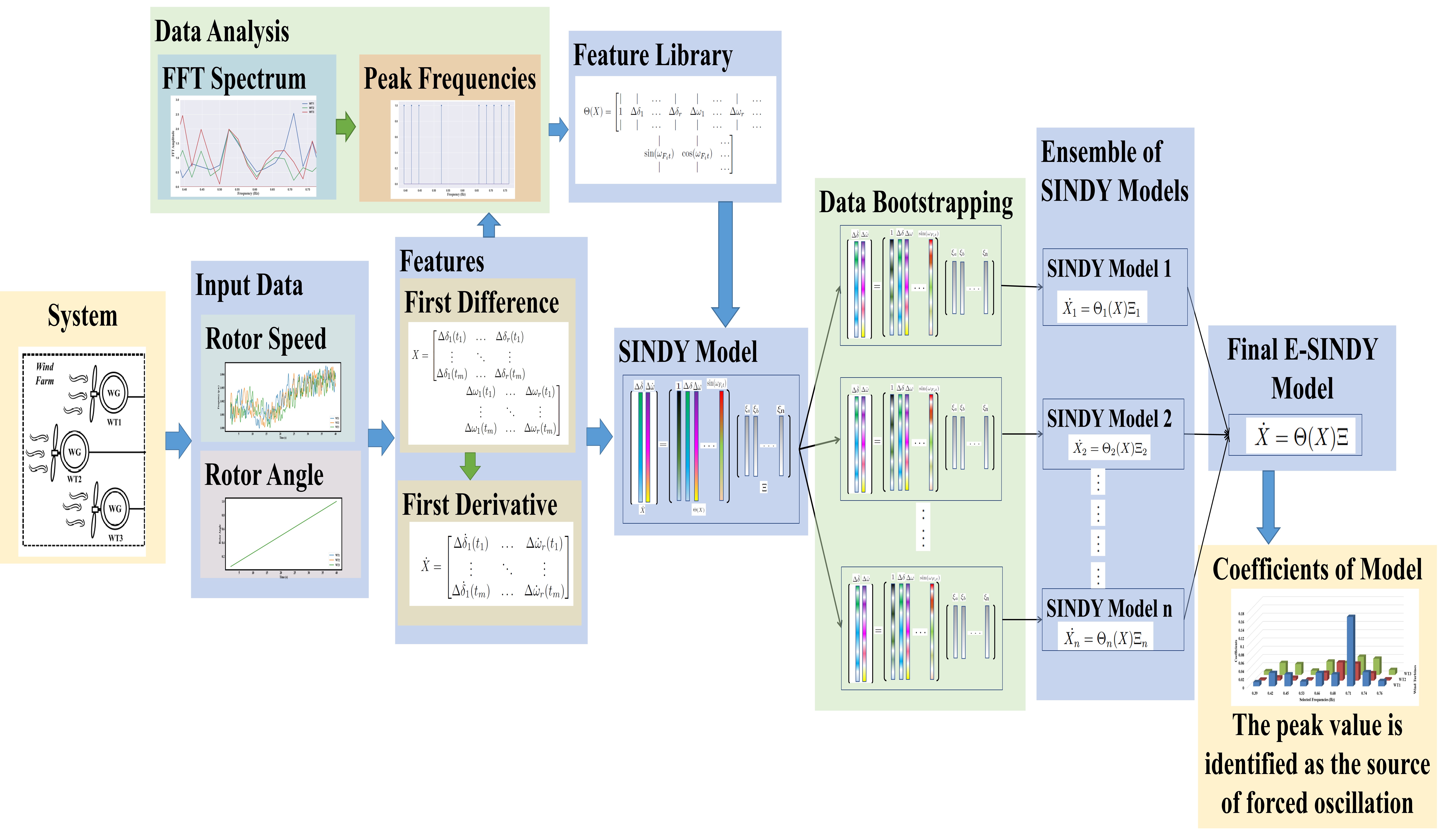}
   \caption{Complete process of the proposed method.}
   \label{fig:Process}
\end{figure*}

\begin{figure}[b]
   \centering
   \includegraphics[width=3.3in]{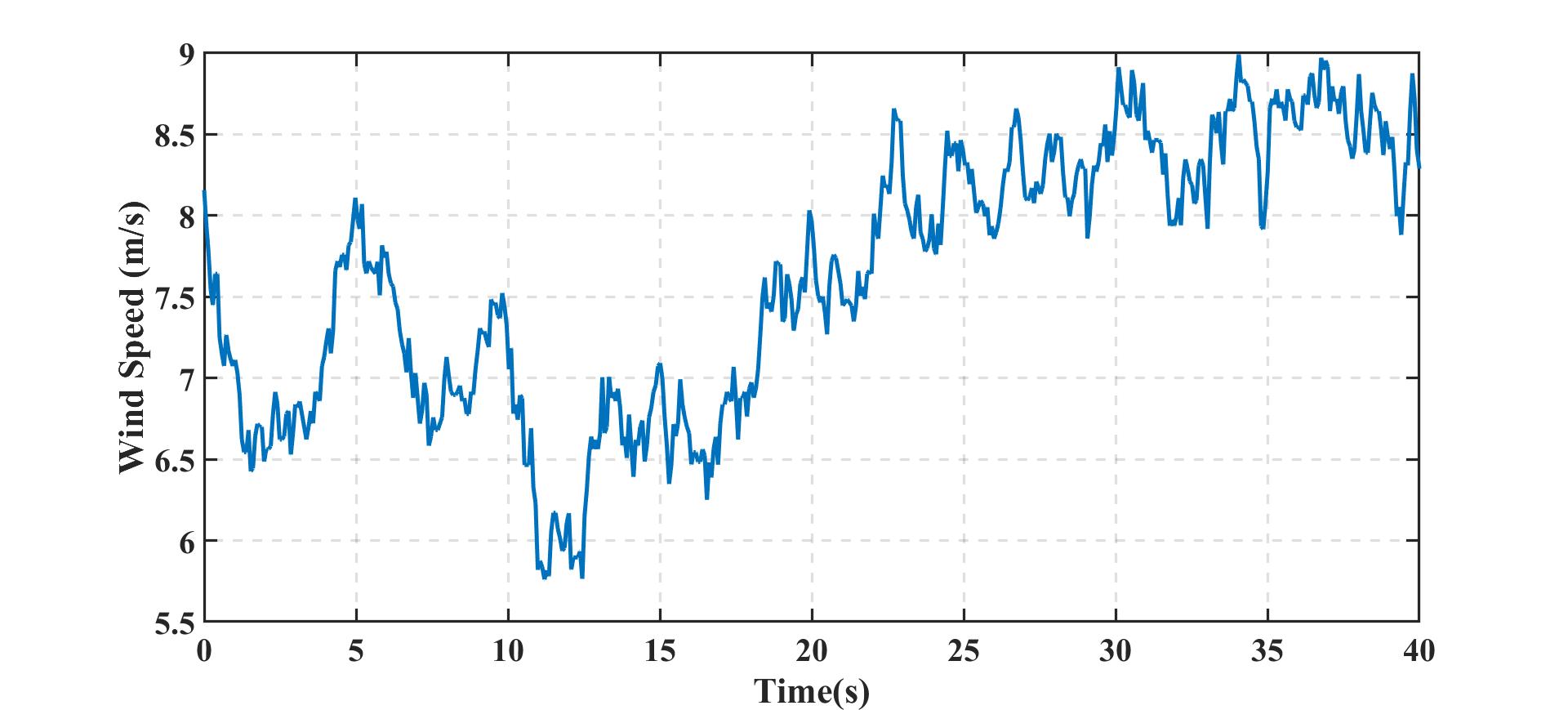}
   \caption{Wind speed at the wind farm.}
   \label{fig:wind_speed}
\end{figure}

To apply the E-SINDy algorithm to wind farm systems, we first simulate the dynamics of a wind farm using the actuator line model \cite{Sorensen2002}. The actuator line model describes the unsteady aerodynamic loads on wind turbine blades based on the local flow conditions, taking into account both the effects of the incoming wind and the wake generated by upstream turbines. The resulting simulation data consists of time series of the velocities and positions of each turbine in the wind farm. We then construct a library of candidate functions that includes all possible nonlinear combinations of the state variables up to a specified polynomial degree. The library is used to generate a matrix of measurements, which is then fed into the E-SINDy algorithm to identify the underlying nonlinear dynamics of the wind farm system. The resulting set of nonlinear functions is used to construct a physically interpretable model of the wind farm dynamics, which aids in designing and optimizing wind farm control strategies. The detailed step-by-step process of the proposed method is shown in Fig. \ref{fig:Process}. The wind speed at the wind farm is shown in Fig. \ref{fig:wind_speed}.

\begin{figure}
\centering
\subfloat[Rotor speed for Case 1.]{\includegraphics[width=0.4\textwidth]{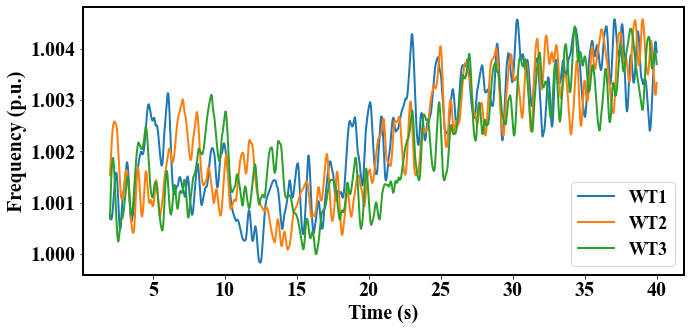} \label{subfig:1_a}}
\quad
\subfloat[FFT analysis for Case 1.]{\includegraphics[height=0.2\textwidth]{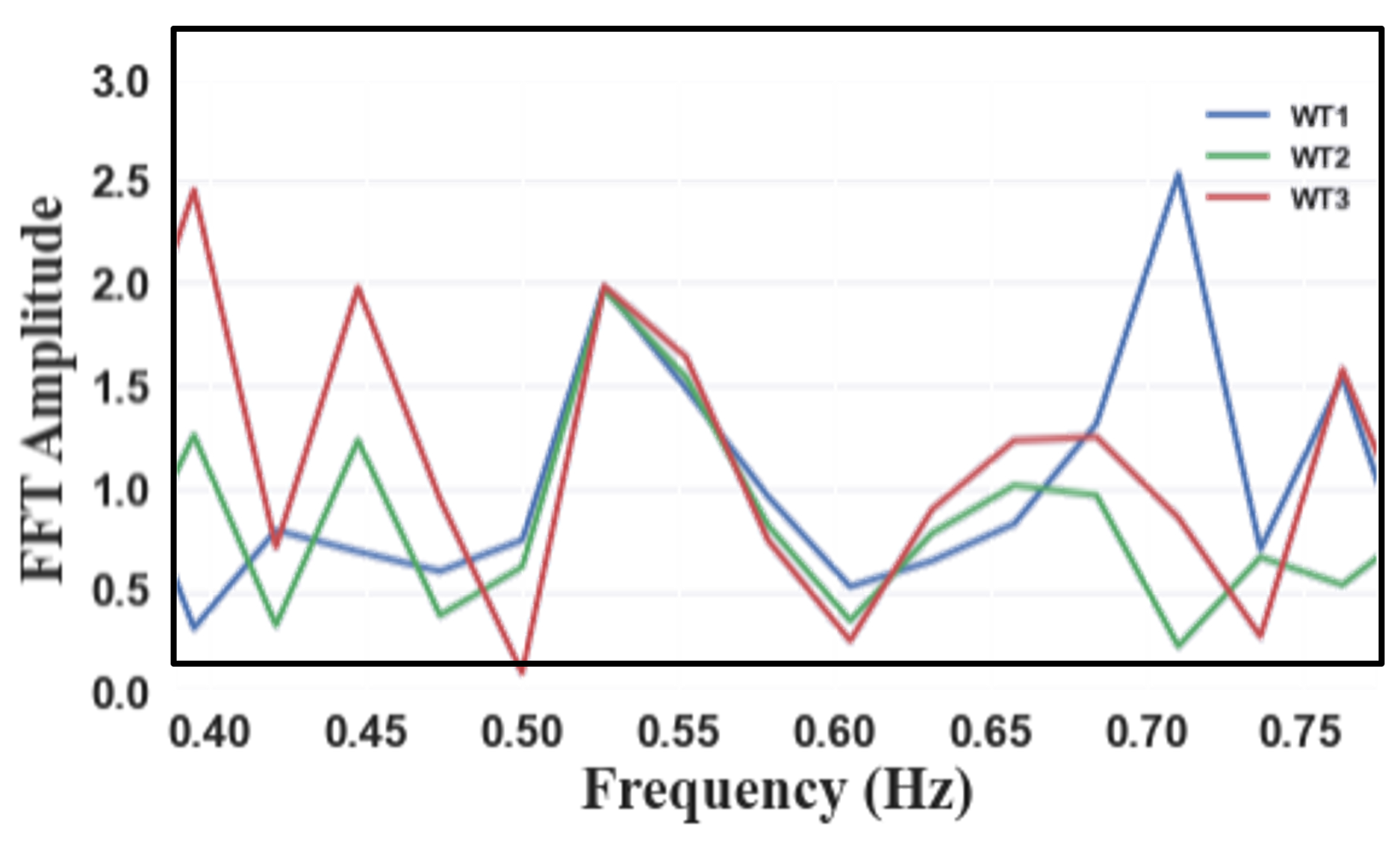} \label{subfig:1_b}}
\quad
\subfloat[Detection of unique frequencies for Case 1.]{\includegraphics[height=0.2\textwidth]{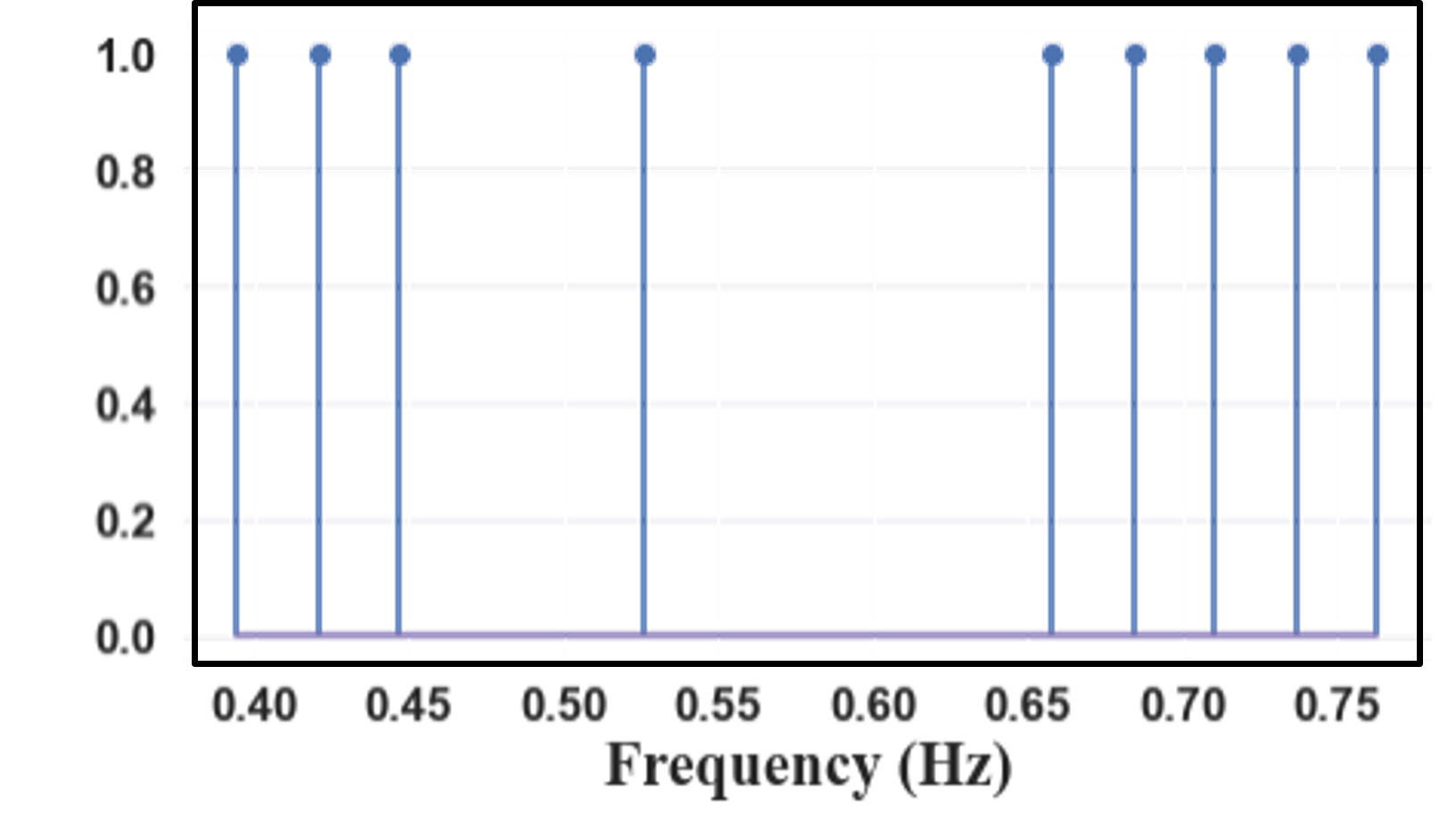} \label{subfig:1_c}}
\caption{The identification of unique frequencies from the rotor speed data of three wind turbines for Case 1.}
\label{fig:case1}
\end{figure}

\begin{figure}[t]
   \centering
   \includegraphics[width=3.3in]{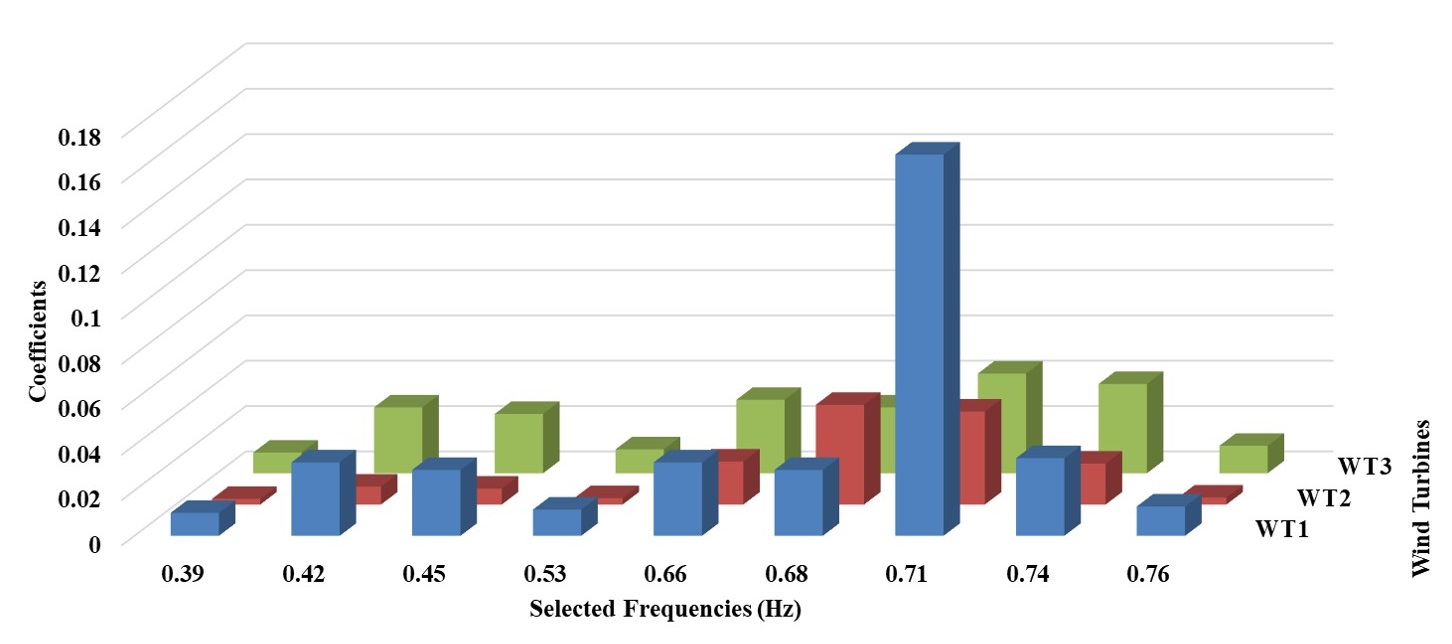}
   \caption{Coefficient matrix terms corresponding to selected frequencies for Case 1.}
   \label{fig:case1_loc}
\end{figure}

\begin{figure}
\centering
\subfloat[Rotor speed  for Case 2.]{\includegraphics[width=0.4\textwidth]{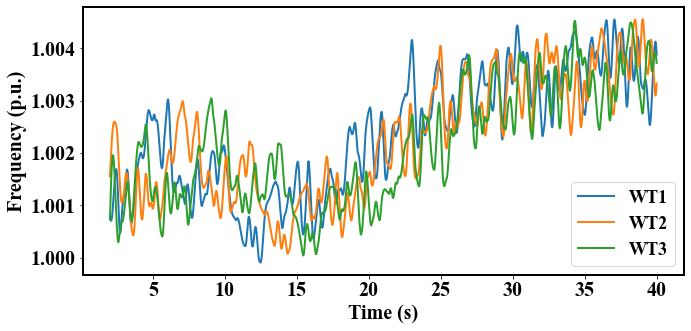} \label{subfig:2_a}}
\quad
\subfloat[FFT analysis for Case 2.]{\includegraphics[height=0.2\textwidth]{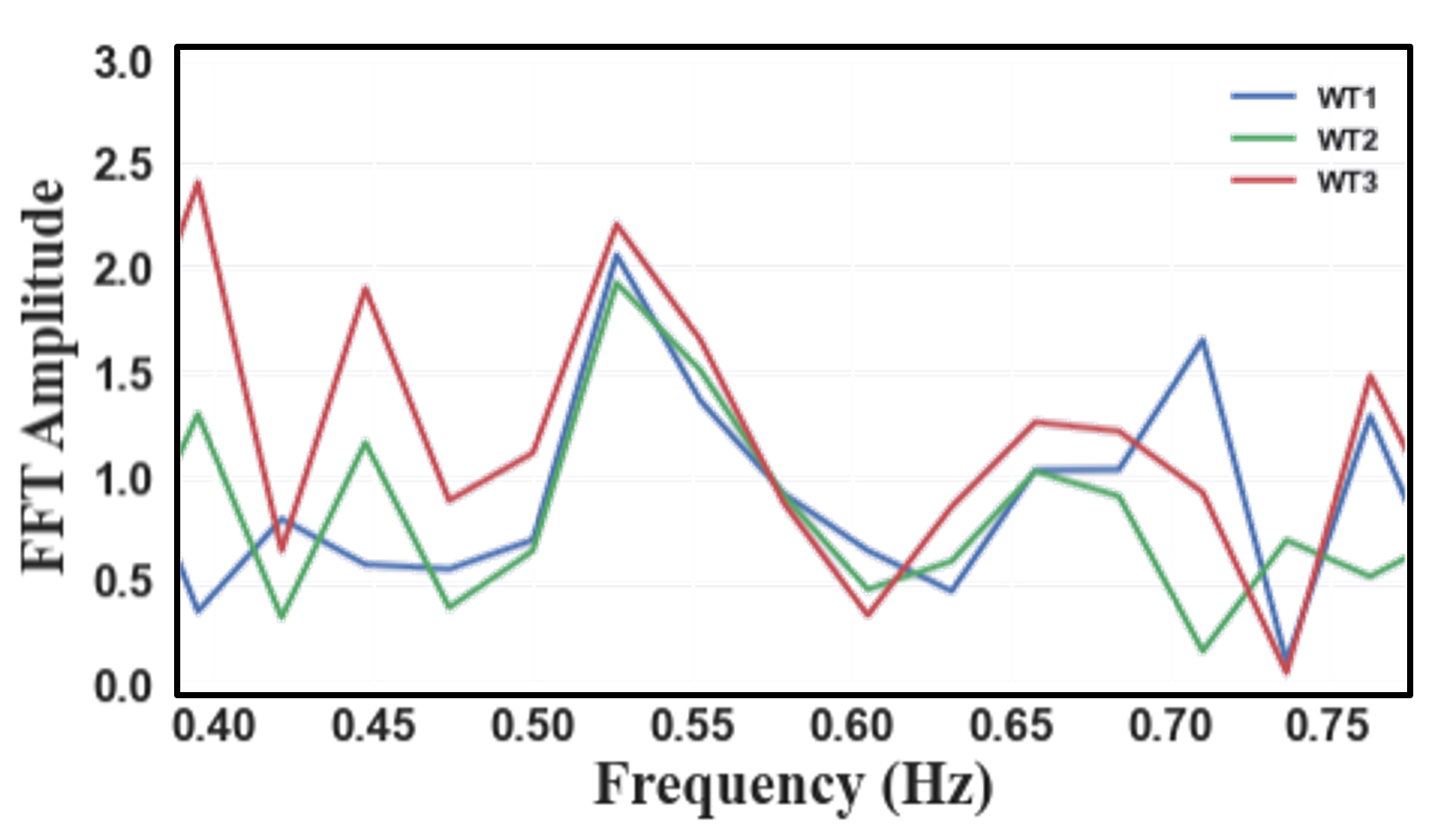} \label{subfig:2_b}}
\quad
\subfloat[Detection of unique frequencies for Case 2.]{\includegraphics[height=0.2\textwidth]{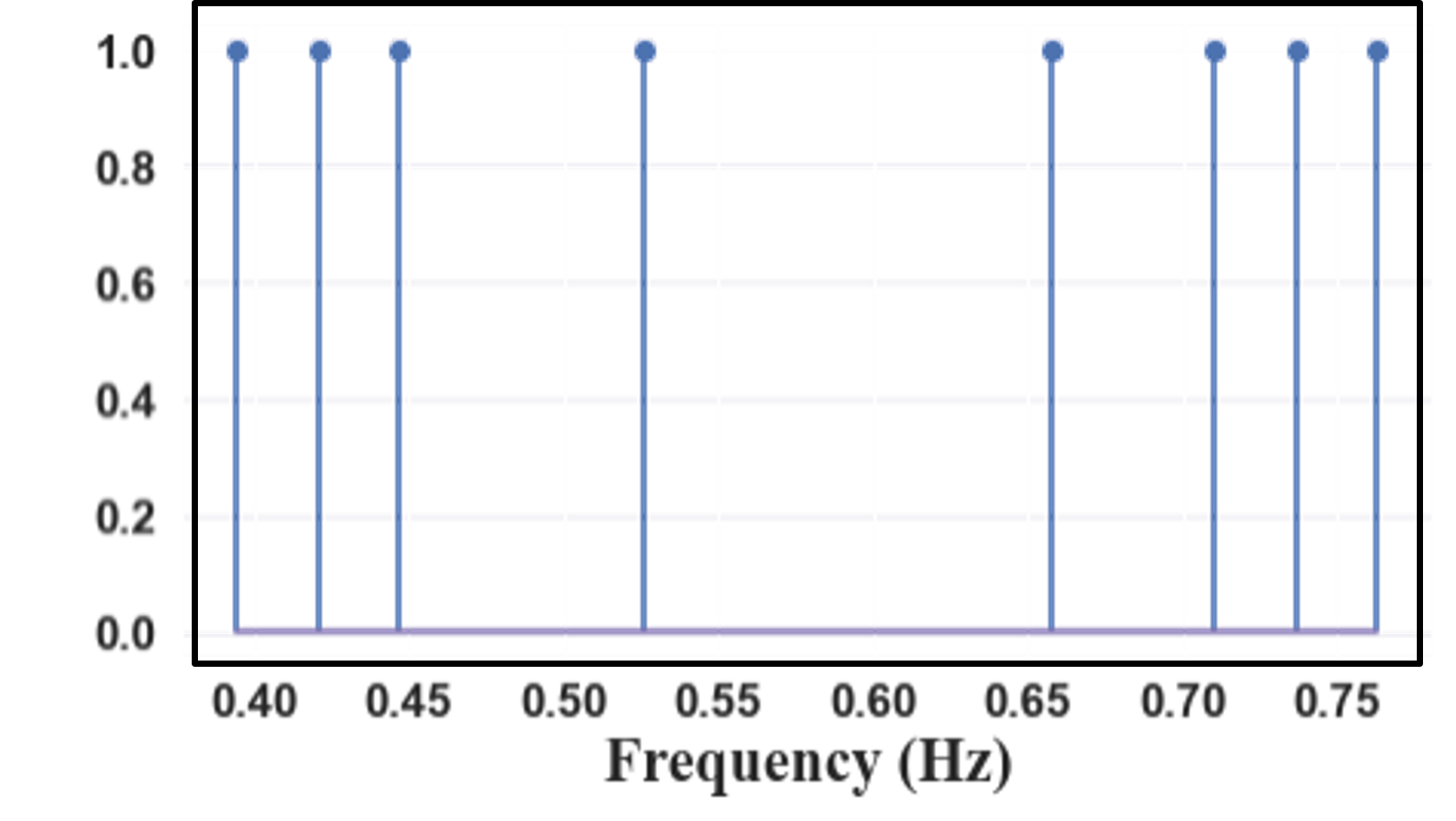} \label{subfig:2_c}}
\caption{The identification of unique frequencies from the rotor speed data of three wind turbines for Case 2}
\label{fig:case2}
\end{figure}

\begin{figure}[t]
   \centering
   \includegraphics[width=3.3in]{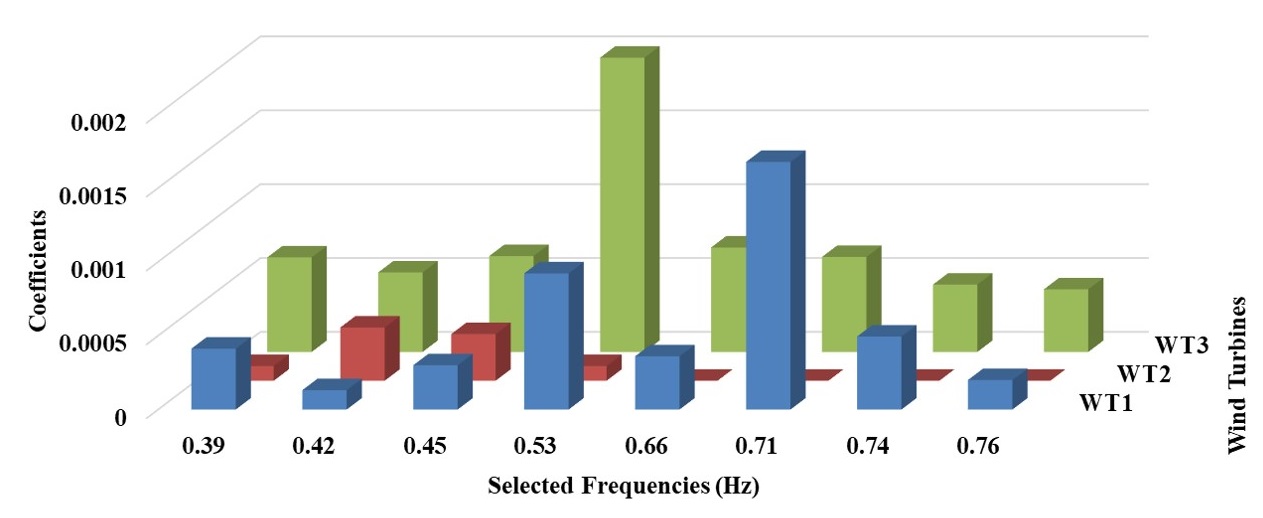}
   \caption{Coefficient matrix terms corresponding to selected frequencies for Case 2.}
   \label{fig:case2_loc}
\end{figure}

\begin{figure}
\centering
\subfloat[Rotor speed for Case 3.]{\includegraphics[width=0.4\textwidth]{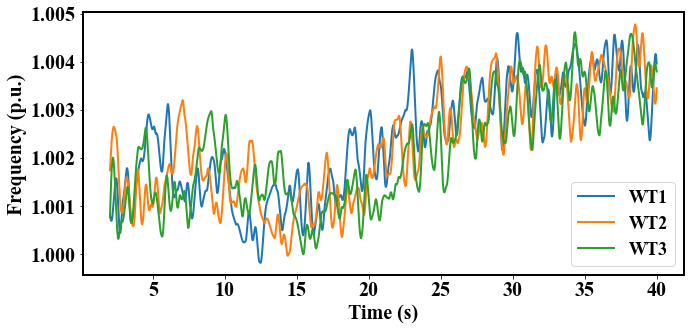} \label{subfig:3_a}}
\quad
\subfloat[FFT analysis for Case 3.]{\includegraphics[height=0.2\textwidth]{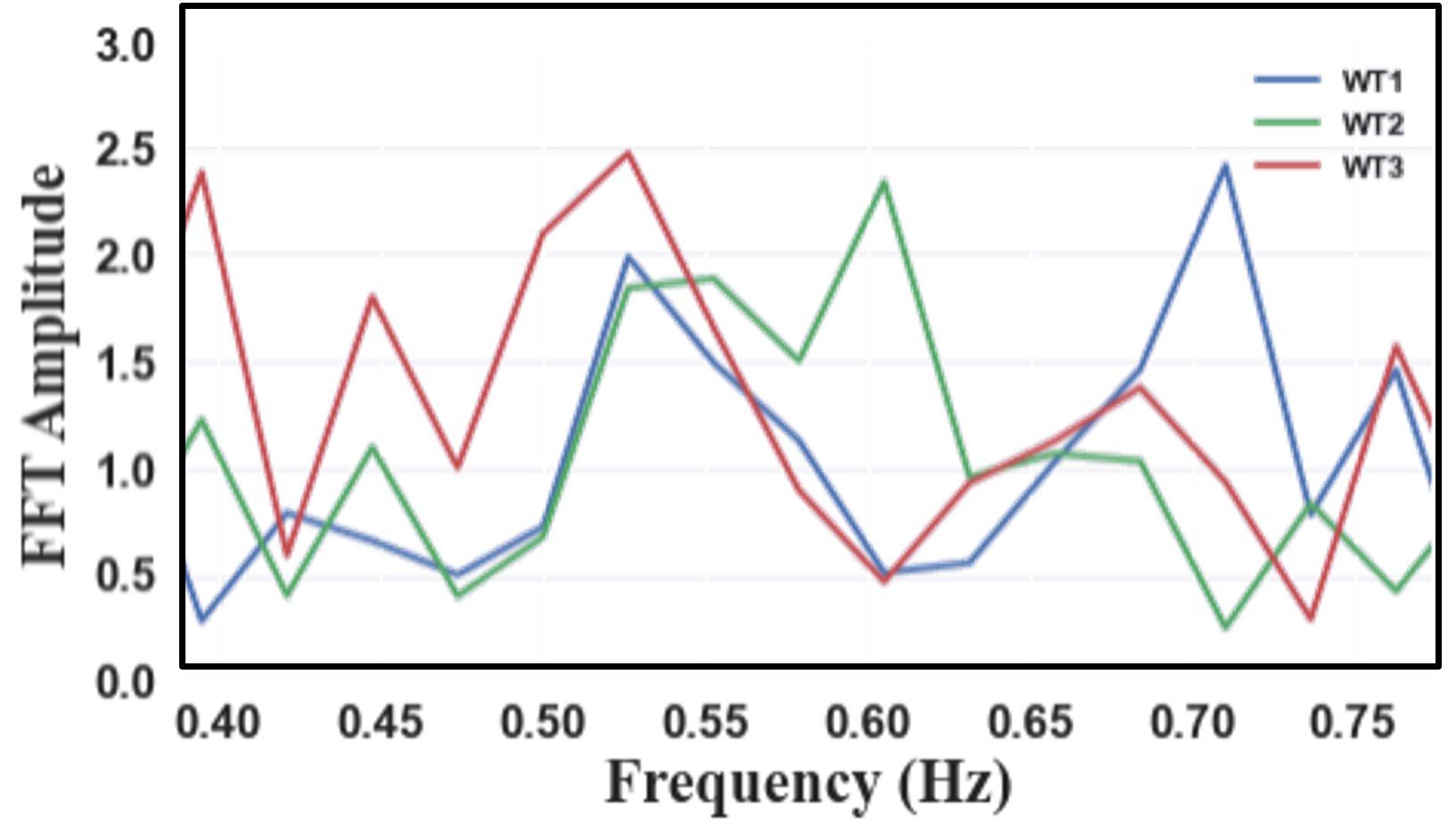} \label{subfig:3_b}}
\quad
\subfloat[Detection of unique frequencies for Case 3.]{\includegraphics[height=0.2\textwidth]{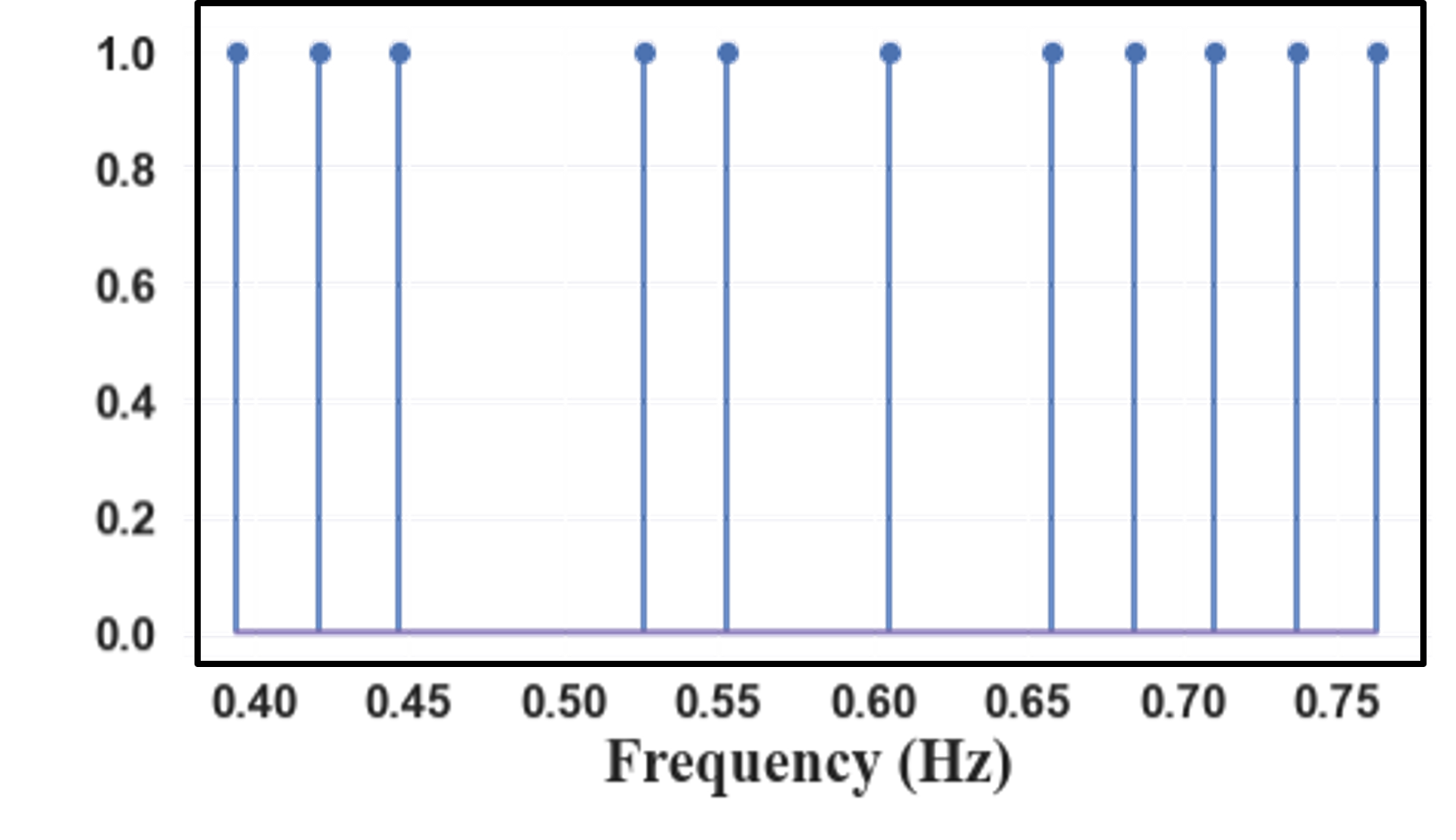} \label{subfig:3_c}}
\caption{ The identification of unique frequencies from the rotor speed data of three wind turbines for Case 3.}
\label{fig:case3}
\end{figure}

\begin{figure}
   \centering
   \includegraphics[width=3.4in]{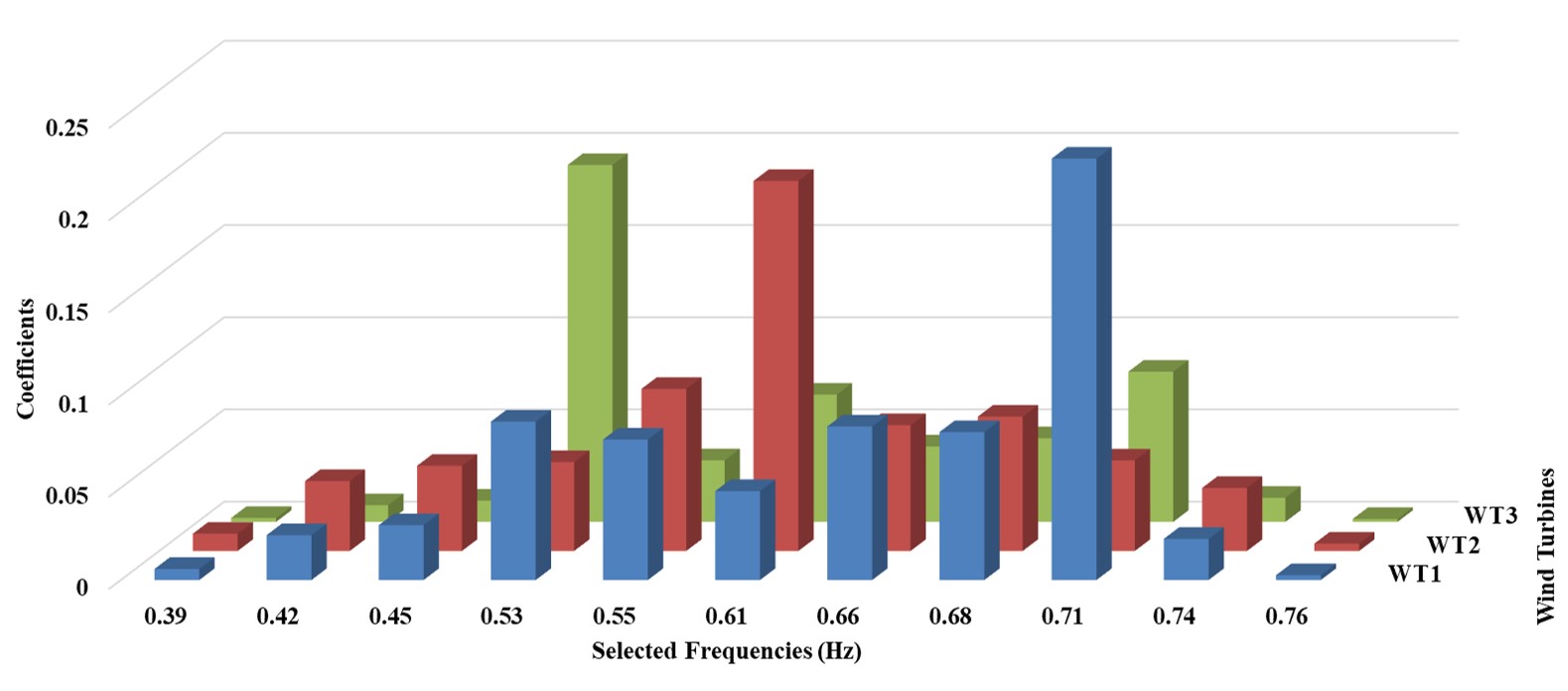}
   \caption{Coefficient matrix terms corresponding to selected frequencies for Case 3.}
   \label{fig:case3_loc}
\end{figure}

We check the performance of the algorithm for all three cases. The trajectory of the angular frequency of all wind turbines is shown in Fig. \ref{fig:case1} \subref{subfig:1_a}, Fig. \ref{fig:case2} \subref{subfig:2_a}, and Fig. \ref{fig:case3} \subref{subfig:3_a}. For all three cases, the trajectories are similar with slight variations. Similarly, in Fig. \ref{fig:case1} \subref{subfig:1_b}, Fig. \ref{fig:case2} \subref{subfig:2_b}, and Fig. \ref{fig:case3} \subref{subfig:3_b}, it is uncertain which wind turbine is the source of forced oscillation for all cases. In the proposed algorithm, we first build the input matrix $X$ and $\dot{X}$. As $\dot{X}$ data is unavailable, we use the finite difference method for estimation. Next, we use FFT and z-score peak-detection methods to obtain the peak candidate frequency terms after removing duplications as shown in Fig. \ref{fig:case1} \subref{subfig:1_c}, Fig. \ref{fig:case2} \subref{subfig:2_c}, and Fig. \ref{fig:case3} \subref{subfig:3_c}. For case 1, the source oscillation frequency is 0.71 Hz and the candidate frequency terms are 0.39, 0.42, 0.45, 0.53, 0.66, 0.68, 0.71, 0.74, 0.76. For case 2, the frequency terms are 0.39, 0.42, 0.45, 0.53, 0.66, 0.71, 0.74, 0.76.  For case 3, the source oscillation frequencies are 0.25 Hz, 0.3 Hz and 0.45 Hz, whereas the frequency terms are  0.39, 0.42, 0.45, 0.53, 0.55, 0.61, 0.66, 0.68, 0.71, 0.74, 0.76. Next, the candidate frequencies are used to construct the feature library $\theta(X)$, and the coefficient matrix $\Xi$ is obtained for the forced oscillation terms. The proposed algorithm calculates the indexes $\Xi$, as demonstrated in Fig. \ref{fig:case1_loc}, \ref{fig:case2_loc}, and \ref{fig:case3_loc}. The source of the oscillations is identified by the peak terms among all the selected frequencies.

If a forced oscillation signal at a frequency close to a resonant frequency appears in the system, it may undergo resonance, which leads to large amplitude oscillations and potentially damage the system. However, whether or not forced oscillations cause resonance in a distribution system also depends on other factors, such as the damping in the system, the strength of the forcing signal, and the overall system design. Consequently, it isn't always true that forced oscillations result in resonance within distribution systems; however, it's essential to assess the possibility of resonance and implement measures to counteract it, such as incorporating damping or modifying the system design to circumvent resonant frequencies. Prompt detection of forced oscillations can aid in their mitigation. Numerous strategies can be employed to mitigate the forced oscillations caused by wind farms connected to a distribution network. These strategies encompass enhancing the control methods of wind turbines to decrease the influence of wind power fluctuations on the system, incorporating energy storage devices to stabilize the wind turbines' output, and augmenting the system's damping by adding shunt reactors or other compensatory devices. Moreover, sophisticated modeling and simulation methods can be utilized to pinpoint potential issues and optimize the wind farm and distribution system's design and operation, minimizing the risk of forced oscillations.

\section{Conclusion}
\label{section:Conclusion}

In conclusion, this paper proposes the use of the ensemble-sparse identification of nonlinear dynamics method for identifying the sources of forced oscillations in wind farms. The approach is proven to accurately identify the nonlinear dynamics of a wind farm model and the sources of forced oscillations in the system through simulations of three distinct scenarios. The suggested algorithm initially constructs the input matrix and employs the FFT and peak-detection methods to acquire the peak candidate frequency terms. Subsequently, the feature library is established using the candidate frequencies. Data bootstrapping is employed to facilitate the ensembling of multiple SINDy models and attain the final model with a precise representation of the system's governing equations. The coefficient matrix of this final model is utilized to identify the forced oscillations. The outliers are then found with the frequencies corresponding to the oscillation sources. The findings indicate that E-SINDy is a valuable instrument for identifying the sources of forced oscillations in wind farms and could assist in devising efficient control strategies to mitigate their detrimental effects.

\bibliographystyle{IEEEtran}
\bibliography{main}

\begin{thebibliography}{10}
\providecommand{\url}[1]{#1}
\csname url@samestyle\endcsname
\providecommand{\newblock}{\relax}
\providecommand{\bibinfo}[2]{#2}
\providecommand{\BIBentrySTDinterwordspacing}{\spaceskip=0pt\relax}
\providecommand{\BIBentryALTinterwordstretchfactor}{4}
\providecommand{\BIBentryALTinterwordspacing}{\spaceskip=\fontdimen2\font plus
\BIBentryALTinterwordstretchfactor\fontdimen3\font minus
  \fontdimen4\font\relax}
\providecommand{\BIBforeignlanguage}[2]{{%
\expandafter\ifx\csname l@#1\endcsname\relax
\typeout{** WARNING: IEEEtran.bst: No hyphenation pattern has been}%
\typeout{** loaded for the language `#1'. Using the pattern for}%
\typeout{** the default language instead.}%
\else
\language=\csname l@#1\endcsname
\fi
#2}}
\providecommand{\BIBdecl}{\relax}
\BIBdecl

\bibitem{trends}
H.~Polinder, J.~A. Ferreira, B.~B. Jensen, A.~B. Abrahamsen, K.~Atallah, and
  R.~A. McMahon, ``Trends in wind turbine generator systems,'' \emph{IEEE
  Journal of emerging and selected topics in power electronics}, vol.~1, no.~3,
  pp. 174--185, 2013.

\bibitem{DD_arxiv}
D.~Dwivedi, K.~Victor Sam Moses~Babu, P.~K. Yemula, P.~Chakraborty, and M.~Pal,
  ``Identification of surface defects on solar pv panels and wind turbine
  blades using attention based deep learning model,'' \emph{arXiv preprint
  arXiv:2211.15374}, 2022.

\bibitem{Wang2016}
J.~Lei, H.~Shi, P.~Jiang, Y.~Tang, and S.~Feng, ``An accurate forced
  oscillation location and participation assessment method for dfig wind
  turbine,'' \emph{IEEE Access}, vol.~7, pp. 130\,505--130\,514, 2019.

\bibitem{Lackner2013}
D.~F. Gayme and A.~Chakrabortty, ``Impact of wind farm placement on inter-area
  oscillations in large power systems,'' in \emph{2012 American Control
  Conference (ACC)}.\hskip 1em plus 0.5em minus 0.4em\relax IEEE, 2012, pp.
  3038--3043.

\bibitem{victor_wind}
P.~S. Kumar, R.~Chandrasena, and K.~V. S.~M. Babu, ``Design and implementation
  of wind turbine emulator using fpga for stand-alone applications,''
  \emph{International Journal of Ambient Energy}, vol.~43, no.~1, pp.
  2397--2409, 2022.

\bibitem{Peherstorfer2020}
B.~Peherstorfer, B.~Kramer, and K.~Willcox, ``Sparse identification of wind
  turbine dynamics: Discovery of sparsity-promoting controls under model
  mismatch,'' \emph{arXiv preprint arXiv:2006.16830}, 2020.

\bibitem{Zhang2021}
W.~Zhang, Y.~Liu, and H.~Yang, ``Reynolds-averaged navier-stokes equations
  based deep learning for aerodynamic design and analysis,'' \emph{Chinese
  Journal of Aeronautics}, vol.~34, pp. 246--257, 2021.

\bibitem{Meng2020}
Y.~Zhu, N.~Zabaras, P.-S. Koutsourelakis, and P.~Perdikaris,
  ``Physics-constrained deep learning for high-dimensional surrogate modeling
  and uncertainty quantification without labeled data,'' \emph{Journal of
  Computational Physics}, vol. 394, pp. 56--81, 2019.

\bibitem{Raissi2018}
M.~Raissi, P.~Perdikaris, and G.~E. Karniadakis, ``Physics-informed neural
  networks: A deep learning framework for solving forward and inverse problems
  involving nonlinear partial differential equations,'' \emph{Journal of
  Computational physics}, vol. 378, pp. 686--707, 2019.

\bibitem{Pawar2021}
B.~Freeman, Y.~Tang, Y.~Huang, and J.~VanZwieten, ``Physics-informed turbulence
  intensity infusion: A new hybrid approach for marine current turbine rotor
  blade fault detection,'' \emph{Ocean Engineering}, vol. 254, p. 111299, 2022.

\bibitem{Jin2019}
G.~Jiang, H.~He, J.~Yan, and P.~Xie, ``Multiscale convolutional neural networks
  for fault diagnosis of wind turbine gearbox,'' \emph{IEEE Transactions on
  Industrial Electronics}, vol.~66, no.~4, pp. 3196--3207, 2018.

\bibitem{Liu2019}
Z.~Liu and L.~Zhang, ``A review of failure modes, condition monitoring and
  fault diagnosis methods for large-scale wind turbine bearings,''
  \emph{Measurement}, vol. 149, p. 107002, 2020.

\bibitem{Pal_Dynamics}
A.~Mandal, Y.~Tiwari, P.~Panigrahi, and M.~Pal, ``Physics aware analytics for
  accurate state prediction of dynamical systems,'' \emph{Chaos, Solitons \&
  Fractals}, 2022.

\bibitem{Sapsis2018}
G.~E. Karniadakis, I.~G. Kevrekidis, L.~Lu, P.~Perdikaris, S.~Wang, and
  L.~Yang, ``Physics-informed machine learning,'' \emph{Nature Reviews
  Physics}, vol.~3, no.~6, pp. 422--440, 2021.

\bibitem{Brunton2016}
S.~L. Brunton, J.~L. Proctor, and J.~N. Kutz, ``Discovering governing equations
  from data by sparse identification of nonlinear dynamical systems,''
  \emph{Proceedings of the National Academy of Sciences}, vol. 113, no.~15, pp.
  3932--3937, 2016.

\bibitem{Rudy2017}
S.~H. Rudy, S.~L. Brunton, J.~L. Proctor, and J.~N. Kutz, ``Data-driven
  discovery of partial differential equations,'' \emph{Science advances},
  vol.~3, no.~4, p. e1602614, 2017.

\bibitem{Loiseau2021}
R.~Rahimilarki, Z.~Gao, A.~Zhang, and R.~Binns, ``Robust neural network fault
  estimation approach for nonlinear dynamic systems with applications to wind
  turbine systems,'' \emph{IEEE Transactions on Industrial Informatics},
  vol.~15, no.~12, pp. 6302--6312, 2019.

\bibitem{ensemble}
U.~Fasel, J.~N. Kutz, B.~W. Brunton, and S.~L. Brunton, ``Ensemble-sindy:
  Robust sparse model discovery in the low-data, high-noise limit, with active
  learning and control,'' \emph{Proceedings of the Royal Society A}, vol. 478,
  no. 2260, p. 20210904, 2022.

\bibitem{Chen2020}
Z.~Ti, X.~W. Deng, and H.~Yang, ``Wake modeling of wind turbines using machine
  learning,'' \emph{Applied Energy}, vol. 257, p. 114025, 2020.

\bibitem{Tibshirani1996}
R.~Tibshirani, ``Regression shrinkage and selection via the lasso,''
  \emph{Journal of the Royal Statistical Society: Series B (Methodological)},
  vol.~58, no.~1, pp. 267--288, 1996.

\bibitem{Montavon2012}
G.~Montavon, K.~Hansen, S.~Fazli, M.~Rupp, F.~Biegler, A.~Ziehe, A.~Tkatchenko,
  O.~A.~v. Lilienfeld, and K.-R. M{\"u}ller, ``Learning invariant
  representations of molecules for atomization energy prediction,'' in
  \emph{Advances in neural information processing systems}, 2012, pp. 530--538.

\bibitem{su_Wind}
C.~Su, W.~Hu, Z.~Chen, and Y.~Hu, ``Mitigation of power system oscillation
  caused by wind power fluctuation,'' \emph{IET Renewable Power Generation},
  vol.~7, no.~6, pp. 639--651, 2013.

\bibitem{Sorensen2002}
R.~J. Stevens, L.~A. Mart{\'\i}nez-Tossas, and C.~Meneveau, ``Comparison of
  wind farm large eddy simulations using actuator disk and actuator line models
  with wind tunnel experiments,'' \emph{Renewable energy}, vol. 116, pp.
  470--478, 2018.

\end{thebibliography}

\end{document}